\documentclass[twocolumn,showpacs,preprintnumbers,draftclsnofoot]{revtex4}
\usepackage{amsmath}
\usepackage{amssymb}
\usepackage{graphicx}
\usepackage{dcolumn}
\usepackage{bm}
\usepackage{amssymb}
\usepackage{graphicx}
\usepackage{dcolumn}
\usepackage{bm}

\begin{document}

\title{ Topological edge states of a graphene zigzag nanoribbon with spontaneous edge magnetism  }
\author{Ma Luo\footnote{Corresponding author:swym231@163.com} }
\affiliation{The State Key Laboratory of Optoelectronic Materials and Technologies \\
School of Physics\\
Sun Yat-Sen University, Guangzhou, 510275, P.R. China}

\begin{abstract}

The topological phases of graphene with spin-orbit coupling, an exchange field, and a staggered-sublattice potential determine the properties of the edge states of the zigzag nanoribbon. In the presence of the Hubbard interaction, the spontaneous magnetization at the zigzag terminations induces sizable magnetic moments at the lattice sites in the bulk region. Thus, the exchange field and staggered-sublattice potential in the bulk region are effectively changed, which in turn change the topological phase. Within a certain parameter regime, quasi-stable excited states of the zigzag nanoribbon exist, which have a different magnetism configuration at the zigzag terminations from the ground state. The quasi-stable excited states could effectively suppress the finite size effect of the topological edge states. The investigation of the topological edge states in the presence of interaction helps the engineering of spintronic nanodevices based on realistic materials.

\end{abstract}

\pacs{0} \maketitle

\section{Introduction}

The topological phases of two-dimensional materials \cite{YuguiYao2011,Motohiko12,YRen16}, such as graphene with a close-proximity substrate \cite{Motohiko13b,Gmitra15,Gmitra16,Zollner16,Cummings17,maluo17,Frank18,maluo19,petra20}, determine the properties of edge states in nanoribbons. The topological phase is dependent on the combination of intrinsic or Rashba spin-orbit coupling (SOC), the antiferromagnetic or ferromagnetic exchange field,  and the staggered-sublattice potential. The quantum spin Hall (QSH) phase supports helical edge states \cite{CLKane05,Zhenhua11}, and the quantum anomalous Hall (QAH) phase supports chiral edge states \cite{Zhenhua10,WangKongTse11,Zhenhua14}. The spin-polarized quantum anomalous Hall (SQAH) phase supports chiral edge states of one spin component \cite{Motohiko13b,petra20}. Topological edge states appear in both armchair and zigzag nanoribbons, which could be harnessed to carry information for spintronic applications \cite{Zutic04,WHan14}.

In the presence of the Hubbard interaction, with the strength designated as $U$, the spin-dependent staggered-sublattice potential is modified at the mean field level. For pristine graphene, the Hubbard interaction induces a phase transition to an antiferromagnetic Mott insulator at $U\approx2.2t$ \cite{Soriano12}, with $t$ being the nearest-neighbor hopping parameter. For realistic graphene-family materials, the strength of the interaction is $U\approx t$ \cite{Schuler13}, which is smaller than the critical value for the Mott insulator transition. For graphene with SOC, an exchange field and a staggered-sublattice potential, our numerical calculation shows that the realistic Hubbard interaction induces a self-energy with a small amplitude. Thus, the phase diagram is hardly changed.

The situation in a zigzag nanoribbon is dramatically changed. The unpaired lattice sites at the zigzag terminations induce sizable spontaneous  magnetization, with opposite directions in the two sublattices \cite{Soriano12,Mitsutaka96,Hikihara03,Yamashiro03,YoungWooSon06,YoungWoo06,Pisani07,Wunsch08,FernandezRossier08,Jung09,Rhim09,Lakshmi09,Jung09a,Yazyev10,Hancock10,Jung10,Manuel10,Feldner11,DavidLuitz11,JeilJung11,Culchac11,Schmidt12,Karimi12,Schmidt13,Golor13,Bhowmick13,FengHuang13,Ilyasov13,Carvalho14,Lado14,MichaelGolor14,PrasadGoli16,Baldwin16,Ortiz16,Hagymasi16,Ozdemir16,Friedman17,ZhengShi17,XiaoLong18}.  The lattice sites at the zigzag terminations are nearly fully magnetized. Thus, the zigzag edge states that carry spin information could be used to construct a spin valve \cite{MinZhou20}. The amplitude of the magnetic moment exponentially decays as the location of the lattice site moves into the bulk region. However, the amplitude does not decay to zero but to a sizable value. The magnetization in the bulk region of the nanoribbon changes the effective exchange field and staggered-sublattice potential, in turn changing the topological phase diagram. The change is more dramatic than that of bulk graphene. The modified phase diagram is numerically studied in this paper by applying the mean field approximation.

For the ground state of the zigzag nanoribbon with the Hubbard interaction, the spontaneous magnetic moments of the two zigzag terminations are antiparallel \cite{Soriano12,Lado14} because the  lattice sites at the two zigzag terminations are in different sublattices. If the topological phase of the bulk region of the nanoribbon is nontrivial, then the topological edge states at the semi-infinite zigzag edges are gapless. However, the finite size effect of a narrow zigzag nanoribbon could gap out the topological edge states because the edge states at opposite zigzag edges couple to each other. Flipping the magnetization of either (both) zigzag termination(s) could drive the systems into quasi-stable excited states. In these states, the wave number of the topological edge states at opposite zigzag edges could be separated into different band valleys such that the coupling between the two edge states is weakened, and the finite size effect is suppressed.

The paper is organized as follows: In Sec. II, the theoretical model and the mean field approximation method are reviewed. In Sec. III, the numerical results of the phase diagrams and the topological edge states are discussed. In Sec. IV, the conclusion is given.

\section{Computational model}

Graphene with proximity-induced SOC and an exchange field is modeled by the tight-binding model with lattice sites on a hexagonal lattice \cite{Frank18,petra20}. In the additional presence of the Hubbard interaction, the Hamiltonian is given as
\begin{eqnarray}
H&=&-t\sum_{\langle i,j\rangle,\sigma}c_{i,\sigma}^{\dag}c_{j,\sigma}+\Delta\sum_{i,\sigma}\eta_{i} c_{i,\sigma}^{\dag}c_{i,\sigma} \nonumber \\ &+&\frac{2i\lambda_{R}}{3}\sum_{\langle i,j\rangle,\sigma,\sigma'}c_{i,\sigma}^{\dag}c_{j,\sigma'}[(\hat{\mathbf{s}}\times\mathbf{d}_{ij})_{z}]_{\sigma,\sigma'} \nonumber \\ &+&\frac{i\lambda_{I}}{3\sqrt{3}}\sum_{\langle\langle i,j\rangle\rangle,\sigma,\sigma'}\nu_{ij}c_{i,\sigma}^{\dag}c_{j,\sigma'}[\hat{\mathbf{s}}_{z}]_{\sigma,\sigma'} \nonumber \\ &+&\sum_{i,\sigma,\sigma'}(\lambda_{AF}\eta_{i}+\lambda_{FM})c_{i,\sigma}^{\dag}c_{i,\sigma'}[\hat{\mathbf{s}}_{z}]_{\sigma,\sigma'} \nonumber \\ &+&\mu\sum_{i,\sigma}\hat{n}_{i,\sigma}+U\sum_{i}\hat{n}_{i,\sigma}\hat{n}_{i,-\sigma} \label{Hamiltonian}
\end{eqnarray}
where $c_{i,\sigma}^{\dag}(c_{i,\sigma})$ is the creation (annihilation) operator for an electron at lattice site $i$ with spin $\sigma$; $\hat{n}_{i,\sigma}$ is the particle number operator in lattice $i$ with spin $\sigma$. The summation with $\langle i,j\rangle$ covers the nearest-neighbor sites, where the hopping strength is assumed to be $t=2.8$ eV. $\Delta$ is the staggered-sublattice potential, with $\eta_{i}=+1(-1)$ for the A (B) sublattice. The summation with $\lambda_{R}$ corresponds to the Rashba SOC, where $\hat{\mathbf{s}}$ is the Pauli matrix vector, and $\mathbf{d}_{ij}$ is the unit vector from lattice site $i$ to $j$. The summation with $\lambda_{I}$ corresponds to the intrinsic SOC, where the summation with $\langle\langle i,j\rangle\rangle$ covers the next-nearest-neighbor lattice sites, and $\nu_{ij}$ is $1$ or $-1$ for the counterclockwise or clockwise hopping path from site $i$ to $j$. The summation with $\lambda_{AF(FM)}$ corresponds to the antiferromagnetic (ferromagnetic) exchange field. $\mu$ is the chemical potential, which is assumed to be zero. With the mean field approximation, the Hubbard interaction is approximated as $U\sum_{i}(\hat{n}_{i,+}\langle\hat{n}_{i,-}\rangle+\hat{n}_{i,-}\langle\hat{n}_{i,+}\rangle)$, where $\langle\hat{n}_{i,\sigma}\rangle$ is the particle number expectation at lattice site $i$ with spin $\sigma$.

The mean field solution is obtained by iterative calculation. The particle number at each site is self-consistently calculated by summing the density profile of the eigenstates in the whole Brillouin zone and, with the weight of the Fermi-Dirac distribution, in each iterative step. If the system preserves the particle-hole symmetry, then the Fermi level is zero. Otherwise, the Fermi level is self-consistently calculated in each step of the iteration. At the first iterative step, the magnetization at each zigzag termination is assumed. Different configurations of the initial magnetization give solutions with a different total energy, which is defined as
\begin{equation}
E_{total}=\sum_{k}\int d\mathbf{k}E_{\mathbf{k},k}-\frac{U}{2}\sum_{i}\langle\hat{n}_{i,+}\rangle\langle\hat{n}_{i,-}\rangle
\end{equation}
where $\mathbf{k}$ is the Bloch wave vector, $k$ is the band index and $i$ is the lattice site index within one unit cell.
The magnetic configuration with the smallest $E_{total}$ is the ground state, and those with larger $E_{total}$ are quasi-stable excited states. In some systems, the iterations with different initial magnetic configurations converge to the same solution, so a quasi-stable excited state does not exist.

By applying the periodic boundary condition with the Bloch phase in a two-dimensional bulk, the self-energy of the bulk is given by the diagonal matrix with matrix elements of $\Sigma_{i,\sigma}=U(\langle\hat{n}_{i,-\sigma}\rangle-1/2)$. The self-energy effectively changes the local potential at each sublattice and spin, in turn changing the effective parameters as follows: $\tilde{\mu}=\mu+\bar{\mu}$, $\tilde{\Delta}=\Delta+\bar{\Delta}$, $\tilde{\lambda}_{AF}=\lambda_{AF}+\bar{\lambda}_{AF}$, and $\tilde{\lambda}_{FM}=\lambda_{FM}+\bar{\lambda}_{FM}$, where
\begin{eqnarray}
\bar{\mu}&=&\frac{1}{4}U(\langle\hat{n}_{A,+}\rangle+\langle\hat{n}_{B,+}\rangle+\langle\hat{n}_{A,-}\rangle+\langle\hat{n}_{B,-}\rangle-2) \nonumber \\
\bar{\Delta}&=&\frac{1}{4}U(\langle\hat{n}_{A,+}\rangle-\langle\hat{n}_{B,+}\rangle+\langle\hat{n}_{A,-}\rangle-\langle\hat{n}_{B,-}\rangle) \nonumber \\
\bar{\lambda}_{AF}&=&\frac{1}{4}U(-\langle\hat{n}_{A,+}\rangle+\langle\hat{n}_{B,+}\rangle+\langle\hat{n}_{A,-}\rangle-\langle\hat{n}_{B,-}\rangle) \nonumber \\
\bar{\lambda}_{FM}&=&\frac{1}{4}U(-\langle\hat{n}_{A,+}\rangle-\langle\hat{n}_{B,+}\rangle+\langle\hat{n}_{A,-}\rangle+\langle\hat{n}_{B,-}\rangle)
\end{eqnarray}
The staggered-sublattice potential and exchange field induce a nonuniform charge distribution in sublattice and spin. In the simple case that only $t$ and $\Delta$ are nonzero and positive, $\langle\hat{n}_{A,\pm}\rangle$ is smaller than $\langle\hat{n}_{B,\pm}\rangle$. As a result, $\bar{\Delta}$ is negative, which effectively decreases the staggered-sublattice potential. In the other case that only $t$ and $\lambda_{AF}$ are nonzero and positive, $\langle\hat{n}_{A,+}\rangle$ and $\langle\hat{n}_{B,-}\rangle$ are smaller than 0.5, while $\langle\hat{n}_{A,-}\rangle$ and $\langle\hat{n}_{B,+}\rangle$ are larger than 0.5, so $\bar{\lambda}_{AF}$ is positive. The antiferromagnetic exchange field is effectively increased \cite{maluo19}. The SOC terms do not induce a sizable nonuniform charge distribution, so they have a small effect on the self-energy. In the model in which all terms in Eq. (\ref{Hamiltonian}) are nonzero, the effective parameter ($\tilde{\Delta}$, $\tilde{\lambda}_{AF}$ or $\tilde{\lambda}_{FM}$) is nearly linearly dependent on the corresponding bare parameter ($\Delta$, $\lambda_{AF}$ or $\lambda_{FM}$, respectively). The topological properties are determined by the effective parameters instead of the bare parameters. With $U=t$, the numerical results show that the effective parameters are only slightly different from the bare parameters, so the phase diagram is hardly modified.

In the zigzag nanoribbon, the Hubbard interaction with $U=t$ induces a spontaneous magnetic moment at the zigzag terminations. The magnetic moment of each lattice site is given by the value $\langle s\rangle^{z}_{i}=\langle n\rangle_{i,+}-\langle n\rangle_{i,-}$. The amplitude of $\langle s\rangle^{z}_{i}$ exponentially decays to a sizable value as the spatial location moves into the bulk region of the nanoribbon. The directions of $\langle s\rangle^{z}_{i}$ on the two sublattices are opposite. The self-energy is spatially dependent, so the effective parameters $\bar{\Delta}$, $\bar{\lambda}_{AF}$, and $\bar{\lambda}_{FM}$ are spatially dependent. The effective parameters in the middle of the nanoribbon determine the topological properties of the bulk region of the nanoribbon, which in turn determine the properties of the edge states. With $U=t$, the effective parameters in the bulk region of the nanoribbon are sizably different from the bare parameters, so the phase diagram is strongly modified.

\section{Numerical results}

\subsection{Zigzag nanoribbon in the QSH or SQAH phase}

\begin{figure}[tbp]
\scalebox{0.63}{\includegraphics{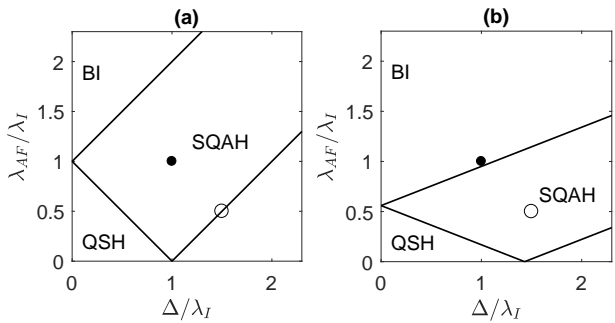}}
\caption{ Phase diagram of the bulk region in the middle of the zigzag nanoribbon with 30 rectangular unit cells (120 atoms) in the width direction, with $U=0$ in (a) and $U=t$ in (b), in the $\lambda_{AF}-\Delta$ parameter regime. The other parameters are $\lambda_{R}=0$, $\lambda_{I}=0.02t$, and $\lambda_{FM}=0$.   }
\label{fig_phaseQSH}
\end{figure}

In the case with intrinsic SOC, a staggered-sublattice potential and an antiferromagnetic exchange field, the phase diagram without the Hubbard interaction is shown in Fig. \ref{fig_phaseQSH}(a). This model could be realized in graphene on an antiferromagnetic monolayer substrate \cite{petra20} or silicene sandwiched between two ferromagnetic substrates with antiparallel magnetization \cite{Motohiko13b}. In the QSH phase, the Chern numbers of the two spins have opposite sign. The $Z_{2}$ topological number is nontrivial, which protects the helical edge state. In the SQAH phase, the Chern number of spin down is one, while that of spin up is zero, so the total Chern number is one. There is only one pair of chiral edge states. In the band insulator (BI) phase, the bulk band is topologically trivial, so there is no topological edge state. In the presence of the Hubbard interaction with $U=t$, the numerical result obtained from the calculation of the zigzag nanoribbon with 30 rectangular unit cells along the width direction shows that the effective staggered-sublattice potential and antiferromagnetic exchange field in the bulk region of the nanoribbon are approximately $\tilde{\Delta}\approx0.7\Delta$ and $\tilde{\lambda}_{AF}\approx1.8\lambda_{AF}+0.005t$. As a result, the phase diagram of the bulk region of the nanoribbon is modified as shown in Fig. \ref{fig_phaseQSH}(b). In the topological phase regime, the properties of the edge states become complicated compared to the corresponding noninteracting model.

\begin{figure*}[tbp]
\scalebox{0.46}{\includegraphics{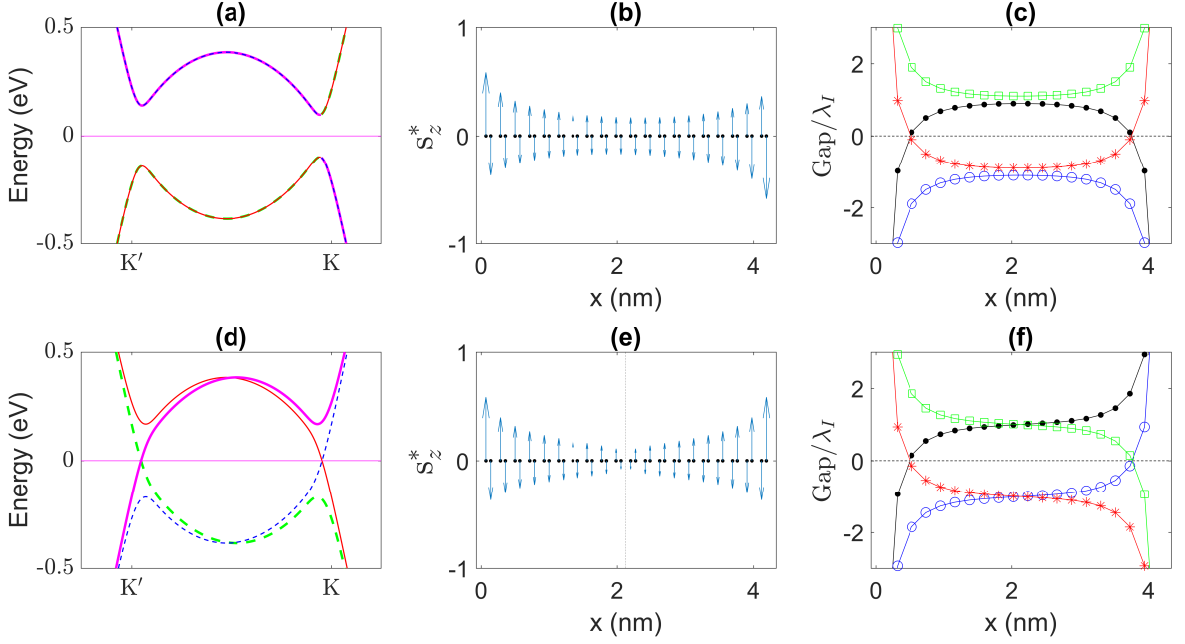}}
\caption{ (a,d) Band structure of the zigzag nanoribbon with 10 rectangular unit cells (40 atoms) in the width direction. The spin up and down bands localized at the left zigzag edge are plotted as blue (dashed) and red (solid) thin lines, respectively. The spin up and down bands localized at the right zigzag edge are plotted as green (dashed) and purple (solid) thick lines, respectively. (b,e) Magnetic moment at each lattice site plotted as arrows with size $s_{z}^{*}=|\langle s\rangle^{z}_{i}|^{0.25}sign[{\langle s\rangle^{z}_{i}}]$ for better visualization. The vertical dashed line in (e) marks the domain wall of antiferromagnetic order. (c,f) Local effective bulk gap of the spin up K valley, spin up K$^\prime$ valley, spin down K valley, and spin down K$^\prime$ valley versus the location along the width direction of the nanoribbon plotted as black (filled dot), blue (empty dot), red (star) and green (empty square) marked lines, respectively. The system parameters are $\Delta=0$, $\lambda_{R}=0$, $\lambda_{I}=0.02t$, $\lambda_{AF}=0$, $\lambda_{FM}=0$, and $U=t$. The results in the top and bottom rows have antiparallel and parallel magnetic configurations of the two zigzag terminations, respectively.   }
\label{fig_bandQSH}
\end{figure*}

For a typical nanoribbon in the QSH phase, the band structures are plotted in Fig. \ref{fig_bandQSH}(a). The magnetic moments of the two sublattices are antiparallel, as shown in Fig. \ref{fig_bandQSH}(b), and have the same configuration as the effective exchange field, i.e., $\langle s\rangle^{z}_{i}$ has the same sign as $-\eta_{i}\tilde{\lambda}_{AF}$. Because the zigzag terminations at the two ends belong to different sublattices, the magnetic moments at the zigzag terminations are antiparallel. For the bulk with the effective parameters, the band gap of spin $\sigma$ in valley $\tau$ is given by $\tilde{\Delta}+\tilde{\lambda}_{AF}\sigma+\tilde{\lambda}_{I}\sigma\tau$, with $\tau=\pm1$ representing the K and K$^{\prime}$ valleys. Since the parameters $\tilde{\Delta}$ and $\tilde{\lambda}_{AF}$ are spatially dependent, the local effective bulk gaps of each spin and valley in the zigzag nanoribbon could be considered as a spatial function along the width direction of the nanoribbon, which is plotted in Fig. \ref{fig_bandQSH}(c). A localized edge state that is either topological or trivial is induced near the location at which the local effective bulk gap flips sign. For spin up, the signs of the local effective bulk gaps of the two valleys are opposite in the bulk region, implying band inversion. Near the two zigzag edges, the local effective bulk gap of the K valley flips sign, while that of the K$^{\prime}$ valley remains open. Thus, the two helical edge states near the two zigzag edges are both in the K valley. The coupling between the two edge states is strong, so the finite size effect gaps out the helical edge states, as shown in Fig. \ref{fig_bandQSH}(a). The first quasi-stable excited state is obtained by flipping the magnetic moment of one of the two zigzag terminations. The band structure, spatial distribution of the magnetic moment, and local effective bulk gaps are plotted in Fig. \ref{fig_bandQSH}(d), (e) and (f), respectively. A domain wall separates the nanoribbon into two regions with opposite antiferromagnetic order, as shown by the dashed line in Fig. \ref{fig_bandQSH}(e). At the domain wall, the parallel magnetic moments of the two nearest-neighbor lattice sites increase the total energy. For spin up, the local effective bulk gap of the K (K$^{\prime}$) valley flips sign near the left (right) zigzag edge. The two helical edge states localized at the left and right zigzag edges are in the K and K$^{\prime}$ valleys, respectively. Thus, the coupling between the two helical edge states is weakened due to the separation in momentum space, so the finite size effect is negligible, as shown in Fig. \ref{fig_bandQSH}(d). The same phenomenon can be found for the spin down helical edge states.

\begin{figure}[tbp]
\scalebox{0.58}{\includegraphics{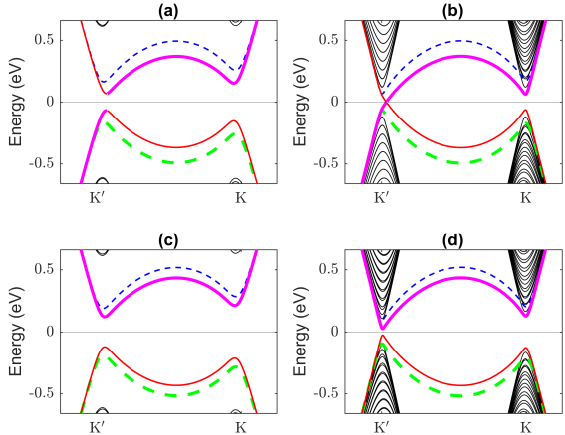}}
\caption{ Band structure of the zigzag nanoribbon with 10 or 60 rectangular unit cells (40 or 240 atoms) in the width direction in (a,c) or (b,d), respectively. The system parameters in (a,b) or (c,d) are given by the empty or filled dot in Fig. \ref{fig_phaseQSH} with $U=t$. The line style of each band localized at the zigzag edge is the same as that in Fig. \ref{fig_bandQSH}(a,d). The solid black lines are the bands of the bulk states.  }
\label{fig_bandSQAH1}
\end{figure}

Since the phase boundary in Fig. \ref{fig_phaseQSH} is significantly changed by the Hubbard interaction, the topological properties of the zigzag nanoribbon are highly dependent on the presence or absence of the Hubbard interaction. For example, the system with $\lambda_{AF}=0.5\lambda_{I}$ and $\Delta=1.5\lambda_{I}$, which is marked as an empty dot in Fig. \ref{fig_phaseQSH}, is at the phase boundary between the SQAH and BI phases in the absence of the Hubbard interaction. In the presence of the Hubbard interaction, the system is in the middle of the SQAH phase regime. The band structure of the ground state of the zigzag nanoribbon with 10 or 60 rectangular unit cells in the width direction is plotted in Fig. \ref{fig_bandSQAH1}(a) or (b), respectively. The band of the spin down localized edge states at the left edge merges into the bulk conduction (valence) band at the K$^{\prime}$ (K) valley, and vice versa for that at the right edge, which characterize the band inversion of the topological phase. In a narrow zigzag nanoribbon, the spin down topological edge states are gapped out by the finite size effect. In a wider zigzag nanoribbon, the finite size effect quickly fades, and the gap is closed. The bands of the bulk states, plotted as solid black lines, are pushed to higher energy due to the finite size effect in a narrow zigzag nanoribbon, as shown in Fig. \ref{fig_bandSQAH1}(a). Another system with $\lambda_{AF}=\lambda_{I}$ and $\Delta=\lambda_{I}$, which is marked as a filled dot in Fig. \ref{fig_phaseQSH}, is in the middle of the SQAH phase regime in the absence of the Hubbard interaction. In the presence of the Hubbard interaction, the system is outside of the SQAH phase regime. The band structures of the zigzag nanoribbon in Fig. \ref{fig_bandSQAH1}(c) and (d) show that increasing the width decreases the gap of the edge states at the K$^{\prime}$ valley. However, the band of the spin down localized edge states at the left (right) edge merges into the bulk valence (conduction) band at both valleys, which characterizes the absence of band inversion. Thus, the edge states are topologically trivial.

\begin{figure*}[tbp]
\scalebox{0.64}{\includegraphics{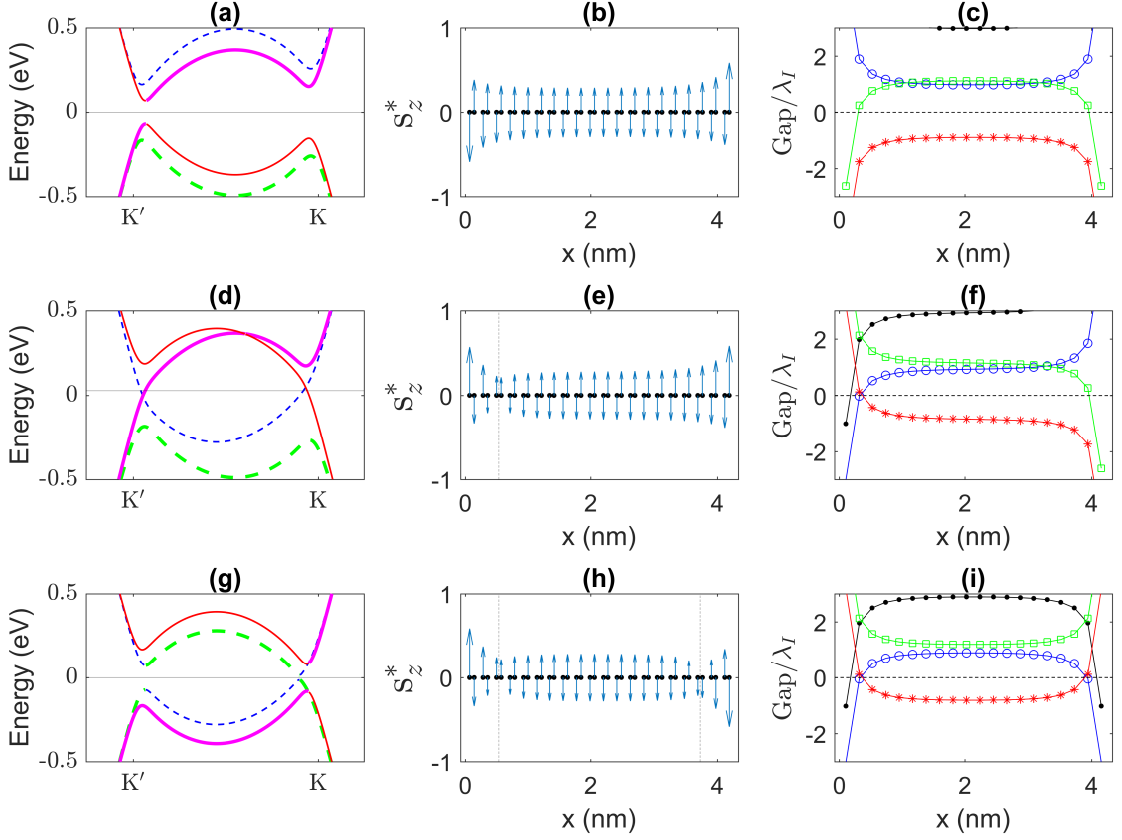}}
\caption{ Same type of plot as in Fig. \ref{fig_bandQSH} for the system marked by the empty dot in Fig. \ref{fig_phaseQSH}, with 10 rectangular unit cells (40 atoms) in the width direction of the zigzag nanoribbon. The top, middle, and bottom rows correspond to the ground, first quasi-stable excited and second quasi-stable excited states, respectively. The system parameters are $\Delta=0.03t$, $\lambda_{R}=0$, $\lambda_{I}=0.02t$, $\lambda_{AF}=0.01t$, $\lambda_{FM}=0$, and $U=t$.   }
\label{fig_bandSQAH2}
\end{figure*}

Similar to the QSH phase, the quasi-stable excited states of the zigzag nanoribbon in the SQAH phase could suppress the finite size effect. As an example, more details about the system with $\lambda_{AF}=0.5\lambda_{I}$ and $\Delta=1.5\lambda_{I}$, which is marked as an empty dot in Fig. \ref{fig_phaseQSH}, are given in Fig. \ref{fig_bandSQAH2}. For the ground state, the configuration of the magnetic moments at the zigzag terminations is antiparallel, with the same configuration as the exchange field, as shown in Fig. \ref{fig_bandSQAH2}(b). As shown in Fig. \ref{fig_bandSQAH2}(c), the local effective bulk gaps of spin up in both valleys are positive in the whole nanoribbon, so the spin up band structure is trivially gapped. The local effective bulk gap of spin down in the K valley is negative in the whole nanoribbon; that in the K$^{\prime}$ valley is positive in the middle of the nanoribbon and then flips sign near the two zigzag edges. Thus, the two topological edge states are both in the K$^{\prime}$ valley, so the finite size effect couples the two topological edge states and gaps out the topological edge bands, as shown by the band structure in Fig. \ref{fig_bandSQAH2}(a). The first quasi-stable excited states are obtained by flipping the magnetic moment of one of the two zigzag terminations. As an example, the quasi-stable excited stated obtained by flipping the magnetic moment of the left zigzag termination has a domain wall of antiferromagnetic order near the left zigzag edge, as shown in Fig. \ref{fig_bandSQAH2}(e). As shown in Fig. \ref{fig_bandSQAH2}(f), the local effective bulk gaps of spin up in both valleys flip sign together near the left edge, so the spin up band remains trivial. The spin up band localized at the left edge accidently crosses the Fermi level without band inversion [blue (dashed) thin line in Fig. \ref{fig_bandSQAH2}(d)]. The local effective bulk gaps of spin down in the K and K$^{\prime}$ valleys flip sign near the opposite zigzag edges, so the finite size effect is negligible. As a result, the two topological edge bands of spin down are gapless [red (solid) thin line and purple (solid) thick line in Fig. \ref{fig_bandSQAH2}(d)]. For the other first quasi-stable excited state, the band structure is the particle-hole inversion of that in Fig. \ref{fig_bandSQAH2}(d), with the same total energy and topological properties. In the second quasi-stable excited state, the magnetic moments of both zigzag terminations are flipped, whose band structure and $\langle s\rangle^{z}_{i}$ are plotted in Fig. \ref{fig_bandSQAH2}(g) and (h), respectively. There are two domain walls of antiferromagnetic order near the two zigzag edges. As shown in Fig. \ref{fig_bandSQAH2}(i), the local effective bulk gaps of spin up in both valleys flip sign together near both zigzag edges. Thus, the spin up bands localized at the left and right edges accidently cross each other at the Fermi level [blue (dashed) thin line and green (dashed) thick line in Fig. \ref{fig_bandSQAH2}(g)]. The local effective bulk gap of the K (K$^{\prime}$) valley has a larger (smaller) amplitude in the bulk region of the nanoribbon, so the finite size effect is smaller (larger), and then, the gap of the band crossing at the K (K$^{\prime}$) valley is smaller (larger). For spin down, the local effective bulk gap of the K valley flips sign near both zigzag edges. Similar to the ground state, the topological edge state in the K valley is gapped out by the finite size effect.

The types of energy levels among the states with varying configurations of the edge magnetization are determined by the interedge superexchange interaction between the magnetic moments on the opposite edges \cite{Jung09a}. For the systems in this subsection, the ground state (first quasi-stable excited state) has an antiparallel (parallel) configuration of the edge magnetization, which has the same types of energy levels as that of the pristine graphene zigzag nanoribbon. Thus, the transition to varying topological phases does not qualitatively change the strength of the interedge superexchange interaction.

\begin{figure}[tbp]
\scalebox{0.58}{\includegraphics{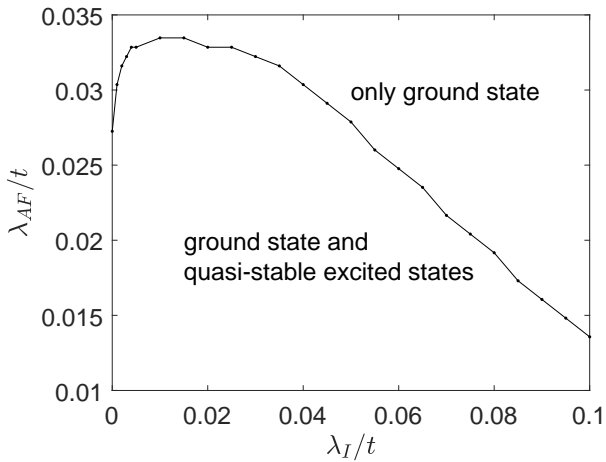}}
\caption{ Phase boundary between the systems with and without the first quasi-stable excited states, in the $\lambda_{AF}-\lambda_{I}$ parameter regime. The other parameters are $\Delta=0$, $\lambda_{R}=0$, $\lambda_{FM}=0$, and $U=t$. The zigzag nanoribbons contain 10 rectangular unit cells (40 atoms) in the width direction. }
\label{fig_phaseFM}
\end{figure}

The numerical results in this section show that the quasi-stable excited states could suppress the finite size effect of the topological edge states in the QSH and SQAH phases. However, in a certain parameter regime, the quasi-stable excited states do not exist. As $\lambda_{AF}$ increases, $\langle s\rangle^{z}_{i}$ in the bulk region has a larger amplitude, which pushes the domain wall of antiferromagnetic order closer to the zigzag termination. When  $\lambda_{AF}$ is larger than a critical value, the domain wall cannot stably exist, so the magnetic moments at the zigzag terminations cannot be flipped from the magnetic configuration of the ground state. The critical value of $\lambda_{AF}$ with varying $\lambda_{I}$ is numerically calculated and plotted in Fig. \ref{fig_phaseFM}. Although $\Delta=0$ is used in the figure, a different value of $\Delta$ does not significantly change the result.

\subsection{Zigzag nanoribbon in the QAH phase}

\begin{figure}[tbp]
\scalebox{0.56}{\includegraphics{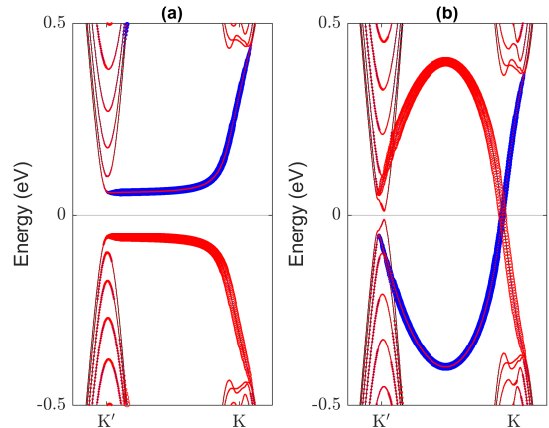}}
\caption{ Band structure of the zigzag nanoribbon with $U=0$ in (a) and $U=t$ in (b). The nanoribbon has 30 rectangular unit cells (120 atoms) in the width direction. The system parameters are $t=2.8$ eV, $\Delta=0.12t$, $\lambda_{R}=0.1t$, $\lambda_{I}=0$, $\lambda_{AF}=0$, and $\lambda_{FM}=0.1t$. The states localized at the left and right zigzag edges are marked by the blue (filled) and red (empty) dots, respectively. The sizes of the blue and red dots represent the level of localization near the left and right zigzag terminations, respectively.  }
\label{fig_bandQAH}
\end{figure}

In the case with Rashba SOC, a staggered-sublattice potential and a ferromagnetic exchange field, the noninteracting system is in the QAH phase if $\lambda_{R}>\Delta$ or in the BI phase if $0<\lambda_{R}<\Delta$ \cite{maluo19a}. In the presence of the Hubbard interaction with $U=t$, the phase boundary is modified as $\lambda_{R}=\tilde{\Delta}$, where $\tilde{\Delta}\approx0.7\Delta$ is obtained from the numerical results. The ferromagnetic exchange field hardly changes, i.e., $\tilde{\lambda}_{FM}\approx\lambda_{FM}$. Since $\tilde{\Delta}$ is smaller than $\Delta$, the system with $\tilde{\Delta}<\lambda_{R}<\Delta$ is driven from the BI phase into the QAH phase by the Hubbard interaction. As an example, the band structure of the zigzag nanoribbon with $\Delta=0.12t$, $\lambda_{R}=0.1t$ and $\lambda_{FM}=0.1t$ is plotted in Fig. \ref{fig_bandQAH}(a) and (b). In the absence of the Hubbard interaction, the nanoribbon is trivial and has no gapless edge band, as shown in Fig. \ref{fig_bandQAH}(a). In the presence of the Hubbard interaction, two pairs of chiral edge states appear because the bulk region is in the QAH phase. For the pair of chiral edge states in the K valley, the bulk gap is large, so the finite size effect is small. For the other pair of chiral edge states in the K$^{\prime}$ valley, the bulk gap is small, so the localization is weak. The finite size effect gaps out the pair of chiral edge states.

\begin{figure}[tbp]
\scalebox{0.45}{\includegraphics{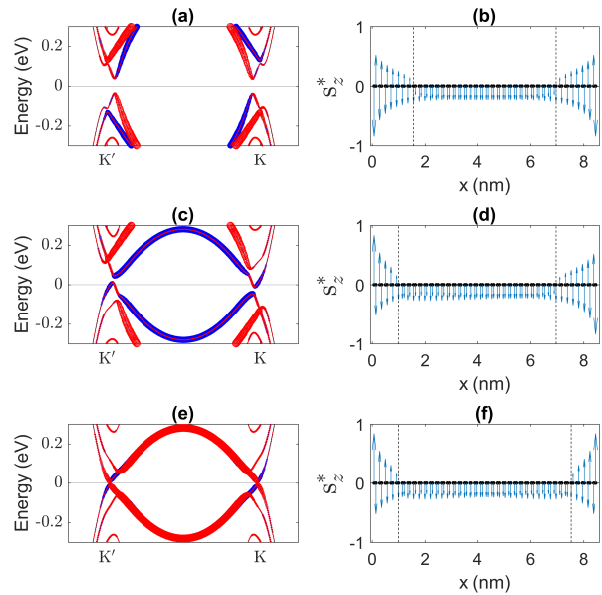}}
\caption{ (a,c,e) Band structure of the zigzag nanoribbon with $\Delta=0$, $\lambda_{R}=0.1t$, $\lambda_{I}=0$, $\lambda_{AF}=0$, $\lambda_{FM}=0.03t$, and $U=t$. The nanoribbon has 20 rectangular unit cells (80 atoms) in the width direction. The states localized at the left and right zigzag edges are marked by the blue (filled) and red (empty) dots, respectively. The sizes of the blue and red dots represent the level of localization near the left and right zigzag terminations, respectively. (b,d,f) Normalized magnetic moment at each lattice site for the corresponding state.  The results of the ground, first quasi-stable excited and second quasi-stable excited states are plotted in (a,b), (c,d) and (e,f), respectively. }
\label{fig_bandQAH1}
\end{figure}

Another example with $\Delta=0$, $\lambda_{R}=0.1t$ and $\lambda_{FM}=0.03t$ is calculated and plotted in Fig. \ref{fig_bandQAH1}. The band structure and magnetic moment distribution of the ground state are plotted in Fig. \ref{fig_bandQAH1}(a) and (b), respectively. In the bulk region, the magnetic moments of the two sublattices are parallel due to the presence of the ferromagnetic exchange field. Near the zigzag terminations, the magnetic moments of the two sublattices are antiparallel due to the spontaneous magnetization of the unpaired lattice sites at the zigzag terminations. The magnetic moments at the zigzag terminations are parallel to the magnetic moment in the bulk region, so the transition between the regions with locally antiparallel and parallel configurations of the magnetic moment is smooth. The domain walls, indicated by the vertical dashed lines in Fig. \ref{fig_bandQAH1}(b), separate the regions with locally parallel and antiparallel configurations of the magnetic moment. Only the region between the two domain walls is in the QAH phase, so the effective width of the topological nanoribbon is decreased. Thus, the finite size effect is enlarged, as shown by the gap of the edge states in Fig. \ref{fig_bandQAH1}(a). In the first quasi-stable excited state, the magnetic moment of one of the two zigzag terminations is flipped, as shown in Fig. \ref{fig_bandQAH1}(d). The transition between the regions with locally antiparallel and parallel configurations of the magnetic moment is steeper, so the total energy is larger. The effective width between the two domain walls is larger, so the finite size effect is weaker, as shown by the small gap of the edge states in Fig. \ref{fig_bandQAH1}(c). In the second quasi-stable excited state, the magnetic moments of both zigzag terminations are flipped, as shown in Fig. \ref{fig_bandQAH1}(f). The finite size effect is further weakened, as shown by the negligible gap of the edge states in Fig. \ref{fig_bandQAH1}(e).

\begin{figure}[tbp]
\scalebox{0.58}{\includegraphics{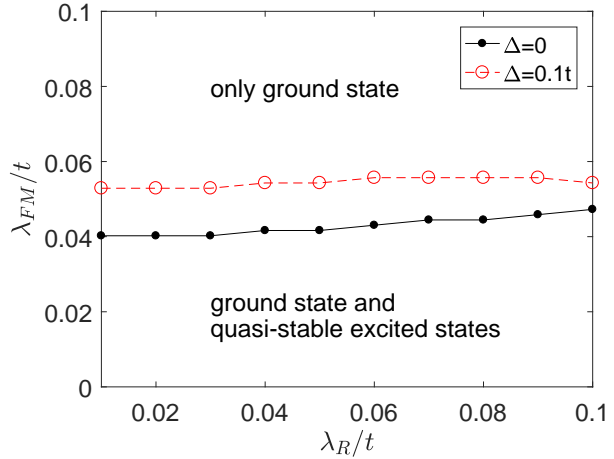}}
\caption{ Phase boundary between the systems with and without the quasi-stable excited states, in the $\lambda_{FM}-\lambda_{R}$ parameter regime, with $\Delta=0$ for the black (solid) line and $\Delta=0.1t$ for the red (dashed) line. The other parameters are $\lambda_{I}=0$, $\lambda_{AF}=0$, and $U=t$. The zigzag nanoribbons contain 20 rectangular unit cells (80 atoms) in the width direction. }
\label{fig_phaseQAH}
\end{figure}

The quasi-stable excited states only exist in the systems with a small ferromagnetic exchange field. As $\lambda_{FM}$ increases, the domain wall in the quasi-stable excited states near the zigzag termination whose magnetic moment is antiparallel to the magnetic moment in the bulk region is pushed toward the zigzag edge. When $\lambda_{FM}$ exceeds a critical value, the domain wall cannot be stable, so the iterative solution converges to the ground state. The phase diagram in the $\lambda_{FM}-\lambda_{R}$ parameter regime with the presence or absence of the quasi-stable excited states is plotted in Fig. \ref{fig_phaseQAH}. A variation in $\lambda_{R}$ slightly changes the critical value of $\lambda_{FM}$. As $\Delta$ become larger, the critical value of $\lambda_{FM}$ increases.

The presence of a ferromagnetic exchange field and Rashba SOC with sufficient strength changes the exchange energy in the interedge superexchange interaction, so the types of energy levels change, i.e., the ground state (first quasi-stable excited state) has a parallel (antiparallel) configuration of the edge magnetization. For the systems with $\Delta=0$, the numerical results show that the boundary between two phase regimes with different types of energy levels is given by $\lambda_{FM}\approx0.0015t-0.038\lambda_{R}$, with $\lambda_{R}<0.04t$. The phase regime with the same types of energy levels as those of the pristine graphene zigzag nanoribbon occupies a small area at the bottom left corner of Fig. \ref{fig_phaseQAH}.

\section{Conclusion}

In conclusion, the Hubbard interaction changes the properties of the topological edge states of the zigzag nanoribbon of graphene. Because the spontaneous magnetic moments at the zigzag terminations significantly change the effective staggered-sublattice potential and antiferromagnetic exchange field, which could in turn change the topological phase in the bulk region, the topological properties of the edge states are modified. The topological edge states of the ground state are gapped out by the finite size effect. The quasi-stable excited states could effectively suppress the finite size effect on the topological edge states. When the amplitude of the exchange field is larger than a critical value, the quasi-stable excited states do not exist.

\begin{acknowledgments}
This project is supported by the National Natural Science Foundation of China (Grant:
11704419).
\end{acknowledgments}

\clearpage

\end{document}